\documentclass[preprint,11pt]{elsarticle}

\makeatletter
\def\ps@pprintTitle{%
 \let\@oddhead\@empty
 \let\@evenhead\@empty
 \def\@oddfoot{}%
 \let\@evenfoot\@oddfoot}
\makeatother

\usepackage[titletoc,title]{appendix}

\makeatletter
  \@addtoreset{chapter}{part}
  \@addtoreset{@ppsaveapp}{part}
\makeatother

\usepackage[latin1]{inputenc}
\usepackage[english]{babel}
\usepackage{caption}

\usepackage{xcolor}
\usepackage{amsmath,amsthm}
\usepackage{relsize}
\usepackage{algorithm}
\usepackage{graphics}
\usepackage{algcompatible}
\usepackage{setspace}
\floatname{algorithm}{}
\let\Algorithm\algorithm
\renewcommand\algorithm[1][]{\Algorithm[#1]\setstretch{1.2}}

\usepackage{euscript}
\usepackage{enumerate}
\usepackage{enumitem}

\usepackage{mathtools}
\DeclarePairedDelimiter\ceil{\lceil}{\rceil}

\begin{document}
\pagestyle{plain}

\begin{frontmatter}

\title{The Baggage Belt Assignment Problem}
\date{}

 \author[1]{David Pisinger \corref{cor1}}
 \author[2]{Rosario Scatamacchia}
\cortext[cor1]{Corresponding author.}
 
  \address[1]{\small Technical University of Denmark, DTU Management Engineering, Akademivej, Building 358, 2800 Kgs. Lyngby, Denmark \\ {\tt dapi@dtu.dk}}
  \address[2]{\small Dipartimento di Ingegneria Gestionale e della Produzione, Politecnico di Torino,\\ Corso Duca degli Abruzzi 24, 10129 Torino, Italy, \\ {\tt rosario.scatamacchia@polito.it}}
  
\begin{abstract}
We consider the problem of assigning flights to baggage belts in the baggage reclaim area of an airport. The problem is originated by a real-life application in Copenhagen airport. The objective is to construct a robust schedule taking passenger and airline preferences into account. We consider a number of business and fairness constraints, avoiding congestions, and ensuring a good passenger flow. Robustness of the solutions is achieved by matching the delivery time with the expected arrival time of passengers, and by adding buffer time between two flights scheduled on the same belt.
We denote this problem as the Baggage Belt Assignment Problem (BBAP). We first derive a general Integer Linear Programming (ILP) formulation for the problem. Then, we propose a Branch-and-Price  (B\&P) algorithm based on a reformulation of the ILP model tackled by Column Generation. Our approach relies on an effective dynamic programming algorithm for handling the pricing problems. We tested the proposed algorithm on a set of real-life data from Copenhagen airport as well as on a set of instances inspired by the real data. Our B\&P scheme outperforms a commercial solver launched on the ILP formulation of the problem and is effective in delivering high quality solutions in limited computational times, making it possible its use in daily operations in medium-sized and large airports.
\end{abstract}
\begin{keyword} Combinatorial optimization \sep baggage belt assignment problem \sep inbound baggage handling \sep branch and price \sep  knapsack problem.
\end{keyword}

\end{frontmatter}


\section{Introduction}

We consider the problem of assigning flights to a fixed set of baggage belts (carousels) for arriving luggage in an airport. The objective is to construct a robust schedule taking passenger and airline preferences into account. We denote this problem as the Baggage Belt Assignment Problem (BBAP). This work is motivated by a real-life application in Copenhagen airport and may as well find application in daily operations of other airports. We focus on the following aspects in the delivery process of bags to passengers. Depending on the arrival gate of an airplane, the passengers may have a quite long walking distance to the baggage claim area. If the passengers, furthermore, have to go through passport control, it can take up to one hour before the passengers can pick up their luggage. If the bags are delivered to the baggage belts before the passengers arrive, it can cause serious congestion problems, since the carousels are filled to the limit and cannot accept further bags. In the light of this, we will consider solutions where the bags are not delivered before the arrival of the passengers at the baggage claim area.
On the other hand, if the passengers arrive much earlier to the baggage claim area than their luggages, then it will lead to dissatisfied passengers, and overcrowded waiting areas. Hence, we assume that every flight has a preferred time for starting the delivery of bags at the baggage claim area. This requested time coincides with the arrival of the passengers at the baggage area. If the bags are delivered later than the requested start time, it will be penalized in the objective function. Also, there may be a maximum delivery start time, namely an upper limit on when bags must be delivered to the passengers. 
For obvious reasons, every flight has a minimum start time for the delivery of the bags. The delivery cannot be scheduled before arrival time to the gate (time on block, TOB) plus the minimum unloading  time and driving time. In our study, we reasonably assume that this delivery time is always smaller than the arrival time of passengers at the belts.

For each flight, we have an expected number of bags. This number can either be reported by the airline, or it can be forecasted from historic information for the same route. Based on the expected number of bags, we can calculate how long time it will take to unload the bags to the baggage belts. This time can be shortened, but it will be strongly penalized in the objective, since it means in practice that two flights will be unloading bags to the same baggage belt. 

Due to an inherent uncertainty in the walking times of the passengers, number of bags to deliver, and productivity of unloading, a robust plan should assign some buffer time between the delivery of bags of two flights. The buffer time is rewarded in the objective function in such a way that the first minutes of buffer time get a large reward, while following minutes of buffer time get a smaller and smaller reward. It is assumed that after a given upper limit on the buffer time, the solution does not become more robust by adding additional buffer time.

Some flights may be incompatible with a given belt. For instance, large flights cannot be assigned to small belts or flights arriving from non-Schengen countries must be assigned to specific belts. For every flight we may have preferences to the chosen baggage belts. Similarly, airline companies may prefer to use baggage belts close to their office, so passengers easily can
solve questions related to damaged/missing bags. Also, if the baggage claim area has several entrances, it will be good to use baggage belts close to the entrance where the passengers will come from.

Finally, we may assume that the schedule should be \emph{fair} in the sense that if a flight $j$ has a requested time for delivery smaller than the time of another flight $j'$, and both flights are scheduled on the same belt, then flight $j$ must be scheduled before flight $j'$. If two flights are scheduled to different belts, then no constraint is imposed on the ordering, since waiting passengers are not standing
at the same belt, and hence do not block for each other.
\bigskip


For an excellent overview of baggage handling at airports we refer to the comprehensive theses by Barth \cite{Barth2013thesis} and Frey \cite{Frey2015}. The Baggage Belt Assignment Problem has been studied in various formulations in
\citep{Barth2013}, \citep{Barth2012a}, \citep{Delonge2012}, and \citep{Frey2012}.
Barth \citep{Barth2013} and Barth and B\"ockmann \citep{Barth2012a} study the assignment of incoming flights to baggage carousels. They model the problem as an extended assignment problem and present a detailed verification study as well as some sensitivity analysis.
Delonge \citep{Delonge2012} presents a model for local and global load balancing of baggage carrousels.
Frey, Kiermaier and Kolisch \cite{Frey2017} consider inbound baggage handling at airports, and present a hybrid heuristic combining a
greedy randomized adaptive search procedure (GRASP) with a guided fast local search (GLS)
and path-relinking.
Pisinger and Rude \cite{Pis2020} report results for the Baggage Belt Assignment Problem using a MIP model that minimizes overlap of aircraft-to-belt assignments, and passenger cross-flow.
A simulation analysis presented in Borille and Correia \cite{Bor13a,Bor13b} investigates how different factors impact the service level of arriving passengers.

However, none of the above papers try to match arrival times of bags to the
arrival times of passengers in order to avoid congestions and ensure a good
passenger flow. 
\bigskip


For the problem we consider in this study, we first derive a general Integer Linear Programming (ILP) formulation. Then, we propose a Branch-and-Price (B\&P) algorithm based on a reformulation of the ILP model tackled by Column Generation (CG) (see, e.g., \cite{DeDeSo05} for an introduction on Column Generation). Our approach relies on an effective dynamic programming algorithm for handling the pricing problems. We tested our solutions methods on a set of real-life data from Copenhagen airport as well as on a set of instances inspired by the real data. Our B\&P scheme appears to strongly outperforms the commercial solver CPLEX 12.9 launched on the ILP model and could be used in real-life applications to improve daily operational processes at the airports. \\

The paper is organized as follows. In Section~\ref{FormDef}, we formally introduce the BBAP and a related ILP formulation. We outline the proposed Branch-and-Price scheme in Section~\ref{BPalgo} and discuss the computational experiments in Section~\ref{CompTest}. Section~\ref{Conc} draws some conclusions. 

\section{Notation and formal problem definition}
\label{FormDef}
In BBAP, let there be given a set of $N = \{1,\ldots,n\}$ flights, a set of $M = \{1,\ldots,m\}$ baggage belts and a set $T$ with $t_{\max}$ start periods $T = \{0,\dots,(t_{\max}-1)\}$. All flights have to be scheduled within the time horizon, i.e., for each flight, the start time of the delivery of the bags plus the corresponding duration on a belt must not exceed $t_{\max}$. Each belt $i \in M$ has associated a set of compatible flights $N_i$ and a productivity $\pi_i$ (e.g., bags per minute) for the unloading operations of the bags. For every flight $j \in N$, let $b_j$ denote the number of bags and let $t_{req}^j$ denote the requested delivery start time corresponding to the arrival of passengers at the baggage area. Thus, for each flight $j$ we have a set of possible start times $T_j = \{t_{req}^j, (t_{req}^j + 1), \dots, (t_{max} - 1) \}$. For every belt $i \in M$ and flight $j \in N_i$ we have a set of possible durations $W_{ij}$. Set $W_{ij}$ includes a nominal duration $w^\prime_{ij} = b_j/\pi_i$ and other durations that may reflect a minimum allowed unloading time, and a maximum time including buffer time. Since the productivity of each belt may differ, the unloading time may be faster at some belts than others. Therefore, the set of possible durations $W_{ij}$ depends on the belt $i$. Let $p_{ijtw}$ be the profit that can obtained by assigning flight $j \in N_i$ to belt $i$ with start time $t$ and duration $w$. The profit reflects preferences to belts and start time and duration. Buffer time, and hence implicitly robustness, is modeled by increasing the profits for longer durations $w$. If a flight has a nominal duration $w'$, then we increase the profit of all durations $w > w'$. Similarly, we penalize durations $w < w'$ to reflect that this may mean that baggages from two different flights will be unloaded at the same time. We also penalize situations where the delivery of the bags of a flight $j$ starts after the requested start time $t_{req}^j$. In these cases, lower profits reflect the presence of disappointed passengers and overcrowded waiting areas. 

Finally, to ensure a fair schedule, we have a set of precedences to consider, namely we must ensure that each belt processes flights in increasing order of the requested start times $t_{req}^j$ $(j= 1, \dots, n)$. From now on,  we assume this ordering of the flights. In order to derive an ILP formulation for the BBAP, we introduce binary variables $x_{ijtw}$ equal to one if and only if flight $j$ is assigned to belt $i$ with start time $t$  and duration $w$. Clearly, the relevant variables $x_{ijtw}$ are the ones with $t + w \leq t_{max}$ and $j \in N_i$. The problem can be formulated as the following ILP model, denoted as $BBAP_{ILP}$.

\begin{align}
\text{$BBAP_{ILP}$:}							 \nonumber \\ 
\max 
&  \sum_{i \in M} \sum_{j \in N} \sum_{t \in T_j} \sum\limits_{\substack{w \in W_{ij}: \\ t + w \leq t_{max}}} 
      p_{ijtw} x_{ijtw} \label{eq:obj} \\
\mbox{s.t.}
&  \sum_{i \in M}  \sum_{t \in T_j} \sum\limits_{\substack{w \in W_{ij}: \\ t + w \leq t_{max}}} x_{ijtw} = 1  
       \qquad  j \in N  \label{eq:assign} \\
&  \sum_{t \in T_j} \sum\limits_{\substack{w \in W_{ij}: \\ t + w \leq t_{max}}} (t+w) \cdot x_{ijtw} \leq  \sum_{t \in T_{j'}} \sum\limits_{\substack{w \in W_{ij'}: \\ t + w \leq t_{max}}} t \cdot x_{ij'tw} + \nonumber \\
& + t_{max}(2 - \sum_{t \in T_j} \sum\limits_{\substack{w \in W_{ij}: \\ t + w \leq t_{max}}} x_{ijtw} - \sum_{t \in T_{j'}} \sum\limits_{\substack{w \in W_{ij'}: \\ t + w \leq t_{max}}} x_{ij'tw})  \nonumber \\
       &  i \in M, j, j' \in N_i ~ (j < j')  \label{eq:overlap} \\ 
&      x_{ijtw} \in \{0,1\} 
       \;\; i \in M, j \in N, t \in T_j, w \in W_{ij}: t + w \leq t_{max} \label{eq:domain}
\end{align}

The objective (\ref{eq:obj}) maximizes the profits of the assignments. The first constraints (\ref{eq:assign}) ensure that every flight $j$ is 
assigned to exactly one belt with a certain duration and start time. Constraints (\ref{eq:overlap}) ensure that if two flights $j,j'$ are both assigned to belt $i$ (i.e. $\sum_{t \in T_j} \sum_{w \in W_{ij}} x_{ijtw} = \sum_{t \in T_{j'}} \sum_{w \in W_{ij'}} x_{ij'tw} = 1$), they will not overlap on the belt and will fulfill the precedence constraints. Otherwise, the constraints are inactive. Finally, constraints (\ref{eq:domain}) define the domain of the decision variables.\\

It is easy to see that the BBAP is strongly NP-hard by a reduction from the bin-packing problem. In the bin-packing problem, we are given a set of items to be packed in bins. Each item has a weight and all bins have the same capacity. The goal is to find the minimum number of bins that permits the packing of all items. Consider the decision version of the problem that asks whether all items can be packed by using at most $K$ bins.
Given an instance of the bin-packing problem, we can construct an instance of the BBAP by associating items with flights and bins with $K$ identical belts. Each belt can process each flight with one possible duration that reflects the weight of the corresponding item. The reduction is completed by setting $t_{\max}$ equal to the capacity of the bins, identical and time invariant profits for all the flights and request start times equal to zero. Solving the BBAP instance and checking if it admits a feasible solution allows us to decide the corresponding instance of the bin-packing problem.          

\section{A Branch-and-Price approach}
\label{BPalgo}

We propose a B\&P approach to effectively tackle the BBAP. The approach is based on a reformulation of model $BBAP_{ILP}$ in which Column Generation is used to solve the LP relaxation. 
The ingredients of the algorithm and related CG framework are described in the following sections. 

\subsection{ILP formulation with fewer constraints, but an exponential number of variables}

We derive an alternative ILP formulation with an exponential number of  variables. Assume that for each belt $i$ we have a set ${\cal S}_i$ of all feasible schedules of the flights. For every schedule $s \in {\cal S}_i$, let $q_{si}$ denote the profit of the schedule, and let the binary parameter $a_{sij}$ be one if schedule $s$ for belt $i$ covers flight $j$. We introduce binary variables $x_{si}$ equal to one if schedule $s$ is used for belt $i$. We obtain the following formulation, denoted as $BBAP_{exp}$.
\begin{align}
\text{$BBAP_{exp}$:}							 \nonumber \\ 
\max 
&  \sum_{i \in M} \sum_{s \in {\cal S}_i} q_{si} x_{si}
      \label{eq:dwobj} \\
\mbox{s.t.} 
&  \sum_{i \in M}  \sum_{s \in {\cal S}_i} a_{sij} x_{si} = 1  
      &  j \in N  \label{eq:dwassign} \\
&  \sum_{s \in {\cal S}_i}  x_{si} \leq 1  
      &  i \in M  \label{eq:dwone} \\
&  x_{si} \in \{0,1\} 
      &  i \in M, s \in {\cal S}_i \label{eq:dwdomain} 
\end{align}

The objective (\ref{eq:dwobj}) maximizes the profits of the selected schedules.
Constraints (\ref{eq:dwassign}) ensure that every flight $j$ is assigned to exactly one belt.
The next constraints (\ref{eq:dwone}) ensure that at most one schedule will be selected
for each belt $i$.
Finally, constraints (\ref{eq:dwdomain}) define the binary domain of the variables.


\subsection{Master Problem}
Since model $BBAP_{exp}$ may contain an exponential number of schedules in sets ${\cal S}_i$, it may be intractable to solve. 
Therefore, we replace the integrality constraints (\ref{eq:dwdomain}) with non-negativity constraints on variables $x_{si}$ and solve the LP-relaxed problem for a subset ${\cal R}_i \subseteq
{\cal S}_i$ of the schedules for each belt $i$. We obtain the following Restricted Master Problem ($RMP$) from model $BBAP_{exp}$.

\begin{align}
\text{$RMP$:}							 \nonumber \\ 
\max 
&  \sum_{i \in M} \sum_{s \in {\cal R}_i} q_{si} x_{si}
      \label{eq:robj} \\
\mbox{s.t.} 
&  \sum_{i \in M}  \sum_{s \in {\cal R}_i} a_{sij} x_{si} = 1  
      &  j \in N  \label{eq:rassign} \\
&  \sum_{s \in {\cal R}_i}  x_{si} \leq 1  
      &  i \in M  \label{eq:rone} \\
&  x_{si} \geq 0 
      &  i \in M, s \in {\cal R}_i \label{eq:rdomain} 
\end{align}

In our algorithmic implementations, we initialize the $RMP$ with a dummy schedule for each belt in order to guarantee the feasibility of the $RMP$ in any node of the B\&P tree. Each dummy schedule processes all flights and has a profit equal to $-\infty$ (i.e., an arbitrarily negative value).

\subsection{Pricing Problem}
\label{thePrice}

Let $y_j$ ($j \in N$) denote the dual variables corresponding to constraints 
(\ref{eq:rassign}), and let $u_i$ ($i \in M$) denote the dual variables corresponding to 
constraints (\ref{eq:rone}). For each belt $i \in M$, we consider the following Pricing Problem $PP_i$ with binary variables $x'_{jtw}$ representing the assignment of flight $j$ to belt $i$ with duration $w$ and start time $t$.
\begin{align}
\text{$PP_i$:}							 \nonumber \\ 
\max
&  \sum_{j \in N_i} \sum_{t \in T_j}  \sum\limits_{\substack{w \in W_{ij}: \\ t + w \leq t_{max}}} 
      (p_{ijtw} - y_j) x'_{jtw} - u_i \label{eq:pobj} \\
\mbox{s.t.}
&  \sum_{t \in T_j}  \sum\limits_{\substack{w \in W_{ij}: \\ t + w \leq t_{max}}} x'_{jtw} \leq 1  
       \qquad  j \in N_i  \label{eq:passign} \\
&  \sum_{t \in T_j}  \sum\limits_{\substack{w \in W_{ij}: \\ t + w \leq t_{max}}} (t+w) \cdot x'_{jtw} \leq 
   \sum_{t \in T_{j'}}  \sum\limits_{\substack{w \in W_{ij'}: \\ t + w \leq t_{max}}} t \cdot x'_{j'tw} + \nonumber \\
& + t_{max}(2 - \sum_{t \in T_j}  \sum\limits_{\substack{w \in W_{ij}: \\ t + w \leq t_{max}}} x'_{jtw} - \sum_{t \in T_{j'}}  \sum\limits_{\substack{w \in W_{ij'}: \\ t + w \leq t_{max}}} x'_{j'tw})  \nonumber \\
       &  
       j, j' \in N_i ~ (j < j')  \label{eq:poverlap} \\ 
&      x'_{jtw} \in \{0,1\} 
       \qquad   j \in N_i, t \in T_j, w \in W_{ij}: t + w \leq t_{max} \label{eq:pdomain} 
\end{align}
The objective (\ref{eq:pobj}) maximizes the reduced cost of the schedule of flights to be determined. Notice that $u_i$ is a fixed contribution in the objective function and can be ignored in the computation of an optimal solution to $PP_i$. Constraints (\ref{eq:passign}) state that each flight $j \in N_i$ could be assigned to belt $i$. 
Constraints (\ref{eq:poverlap}),
in combination with constraints (\ref{eq:passign}), 
ensure that two flights $j,j' \in N_i$ 
do not overlap if they are processed on belt $i$ (as constraints  \eqref{eq:overlap}). Also, these constraints ensure the precedence constraints.\\

It is easy to see that $PP_i$ contains the 0-1 Knapsack Problem (see monographs \cite{KePfPi04}, \cite{MarTot90}) as special 
case, using the same arguments as for BBAP. The input size of $PP_i$ is 
$\Theta(n w_{\max} t_{\max})$ if the parameters are given explicitly, where $w_{\max}$ denotes the maximum size of an interval $W_{ij}$. If instead  the parameters can be calculated ad-hoc when needed, the order of magnitude of the number of parameters is $\Theta(\log_2 w_{\max} t_{\max})$ for each flight, thus reducing the input size to only $\Theta(n\log_2 w_{\max} +n\log_2 t_{\max})$.\\
 
At each node of the Branch-and-Price tree, we solve the $RMP$ and obtain the corresponding values of dual variables $y_j, u_i$. Then, we solve to optimality the pricing problems $PP_i$ for every belt $i \in M$. If a schedule for belt $i$ with positive reduced cost is found, the schedule is added to set ${\cal R}_i$. Then, we solve the $RMP$ again and the CG process is repeated until no schedules with positive reduced costs can be determined. An optimal solution of the $RMP$ provides an Upper Bound ($UB$) on the BBAP. Also, we keep track of integer solutions of the $RMP$ along the CG iterations to possibly update the current Lower Bound ($LB$), namely the best feasible solution computed so far.\\

Clearly, the pricing problems should be solved effectively to speed up the convergence of the iterative process. Here, we notice that each pricing problem $PP_i$ can be seen as a variant of the Knapsack Problem where the position of the packed items impacts the profits and there are precedence constraints among the items in addition to the standard capacity constraint. We denote this problem as the \emph{Position Dependant Knapsack Problem} (PDKP). A crucial observation is that, in each problem $PP_i$, we have an implicit capacity constraint because two flights cannot overlap on a belt and the flights must be scheduled within the specified time horizon. Since we assume that flights (items) in $N_i$ are sorted by increasing requested start times, we design a recursion that implicitly satisfies the non-overlap requirement and iteratively processes flights according to the specified order to satisfy the precedence constraints. This means that we only have to choose flights and their duration, considering the flights in $N_i$, up to (capacity) $t_{\max}$. This leads to an effective dynamic programming algorithm to solve the $PP_i$ related to a belt $i$.\\

In the dynamic program, let $f(t,j)$ be the maximum profit we can obtain by considering only the first $j$ flights in $N_i$ and a time horizon of $t$. For each $j = 1, \dots, |N_i|$ we indicate by $[j]$ the associated flight in set $N$. As initial conditions, we set $f(0,j) = 0$ for $j = 1, \dots, |N_i|$ and $f(t,0) = 0$ for $t = 0, \dots, t_{\max}$. We also assume $f(t,j) = -\infty$ if $t < 0$ or $j < 0$.  We then have the following recursion within two nested for-loops:
\begin{align}
& \forall j = 1,\dots, |N_i|, \; \forall t = 1, \dots, t_{\max}: \nonumber\\
\label{DPrec}
& f(t,j) = \max \left\{ \begin{array}{lll}
                                  f(t, j-1),\\
                                  f(t-1, j), \\
                                  \max\limits_{w \in W_{i[j]}} \{f(t - w, j-1) + (p_{i[j](t-w)w} - y_{[j]}) \mid t_{req}^{[j]} \leq t - w  \}.       
                                  \end{array} \right\}  
\end{align}
 
The term $f(t, j-1)$ in the recursion represents the case where flight $j$ is not selected. The inner maximization expression considers the selection of flight $j$ with different durations on belt $i$ and finish time $t$. Clearly, the relevant cases here are only the ones with a start time $(t-w)$ larger than (or equal to) the requested start time $t_{req}^{[j]}$. 
The term $f(t-1, j)$ indicates a possible update from the subproblem with the first $j$ flights and the immediately preceding time $(t-1)$. Notice that this term is needed to guarantee the correctness of the recursion for any arbitrary distribution of the profits, as a-priori we do not know if the schedule of flight $j$ should finish in time $t$ in an optimal solution of subproblem $f(t,j)$.  \\

The time complexity of the recursion  for any pricing problem is $O(nw_{\max} t_{\max})$. 
The space complexity of the recursion is $O(n t_{\max})$. Notice that we have $w_{\max} \ll t_{\max}$ in realistic BBAP instances (see Section \ref{CompTest}). The time complexity of $PP_i$ is linear in the input size $\Theta(n w_{\max} t_{\max})$ if the parameters are given explicitly, but pseudo-polynomial if the parameters can be calculated ad-hoc.\\

The optimal solution value is given by $f(t_{\max},|N_i|)$. Correspondingly, the maximum reduced cost in problem $PP_i$ is equal to $f(t_{\max},|N_i|) -  u_i$. Without going into details, we point out that an optimal schedule of a given pricing problem, i.e. a new column for the $RMP$, can be recovered from the DP entries by implementing an appropriate backtracking procedure and using auxiliary data structures (without increasing the space and time complexities).\\

\medskip 

\subsection{Branching strategy}
In our  B\&P approach we adopt a branching strategy where at most $m$ children are created from a father node by assigning a flight to each compatible belt. 
We branch on 
the flight that appears more times in the fractional columns of an optimal solution of the $RMP$ in the father node. Clearly, if no fractional columns exists, 
the father node can be pruned. \\ 
Else, each child node inherits all the columns of the father node that are not forbidden by the branching decisions. Correspondingly, the $RMP$ is solved by taking into account the previous branching operations in the pricing problems. More precisely, when we solve the pricing problem for a given belt, we artificially increase all profits of the flights that must be assigned to the belt, namely we add a sufficiently large value to the profits of the flights to ensure their selection by the dynamic program depicted in Section \ref{thePrice}. Notice that if the time horizon is not large enough to process all flights assigned to the belt, we can prune the node by infeasibility. Likewise, in the pricing problem of a belt we do not consider the flights that must be assigned to other belts. Finally, we adopt a Best First Search strategy in the exploration of the B\&P tree, namely we first select the node with the largest $UB$.

\section{Computational experiments}
\label{CompTest}
We tested the B\&P algorithm on a set of real-life data from Copenhagen airport as well as on a set of randomly generated instances inspired by the real data. We compared the proposed approach with the commercial solver CPLEX 12.9 launched on the ILP formulation of the problem. We first provide a description of the features of the instances and then discuss the performance of our approach. 

\subsection{Instances and experimental settings}

We first consider real data from Copenhagen airport on five representative days. For each day, we considered time slots of two hours (00:00-02:00, 02:00-04:00, $\dots$, 22:00-24:00) and corresponding instances with $t_{\max} = 120$ minutes. The number of flights was 22 on average with up to 44 in the peak hours. Copenhagen airport normally has 7 belts but at the considered days 2 belts were closed due to infrastructure changes, leaving only 5 belts open.
All flights could be assigned to 5 claim belts (i.e., $N_i  =  N$ for each belt $i$). The belts had the same productivity $\pi_i$ (unloading speed) of 10 bags per minute. Besides, the unloading speed of the last belt could be doubled for flights with 100 bags or more, since it has two unloading stations. We derived the requested delivery time $t_{req}^j$ for each flight $j$ by considering the arrival time of the flight and the time needed for passengers to get to the baggage claim area. For each belt $i$ and flight $j$, we considered a nominal duration (in minutes) equal to the ratio, rounded up to the nearest integer value, between the number of bags and the productivity of the belt, i.e., $w^\prime_{ij} = \ceil*{b_j/\pi_i}$. According to the value of $w^\prime_{ij}$, we defined a set of durations $W_{ij}$ with five duration values (including $w^\prime_{ij}$) and at most two positive duration values smaller than $w^\prime_{ij}$ (so there are at least two durations larger than  $w^\prime_{ij}$ to represent buffer times). Two consecutive durations have a difference of two minutes. Profits $p_{ijtw}$ were generated taking into account both start time $t$ and duration $w$. 
Due to lack of information, we did not consider in the profits possible preferences of air companies for specific belts. More precisely, each profit $p_{ijtw}$ is set to the rounded value (to the nearest integer) of the following weighted sum 
$$\alpha f(w) + (1-\alpha) g(t),$$
with $0 < \alpha < 1$, $f(w) = \beta_1 \frac{e^{(w - w^\prime_{ij})}}{1 + e^{(w - w^\prime_{ij})}}$, $g(t) = \beta_2 \frac{t_{max} - t}{t_{max} - t_{req}^j}$. The sigmoid function $f(w)$ is used to model decreasing durations and increasing buffer times. Function $g(t)$ contributes to decreasing profits with the increase of start time $t$: Starting from a value of $\beta_2$ for $t =t_{req}^j$, the profits are linearly decreased towards 0 as the value of $t$ gets closer to $t_{max}$. In our tests, we consider  parameters $\beta_1 = 500, \beta_2 = 500$, $\alpha = 0.5$ and $\alpha = 0.8$ to induce more robust solutions by increasing the contributions of buffer times in the objective function. We considered a total number of 78 instances related to the data from Copenhagen airport.\\

To get a larger test-bed, we also generated further instances inspired by the real-life data from Copenhagen airport. We considered instances with different number of flights/belts, i.e., $n=30$/$m= 5$ and $n=50$/$m=10$ and with $t_{max} = 120$ minutes. We focused on instances where each belt can schedule each flight as these instances are expected to be more difficult to solve.  For each belt $i$, we generated its productivity $\pi_i$ (baggages per minute) uniformly random in $[10, 20]$.
We generated for each flight $j$ a number of bags $b_j$ uniformly random in $[50,300]$ and a requested start time $t_{req}^j$ uniformly random in $[0, \frac{t_{max}}{2}]$ and in $[0, \frac{3t_{max}}{4}]$ to evaluate different congestion scenarios of the baggage claim area.  Nominal durations $w^\prime_{ij}$, sets $W_{ij}$ and profits $p_{ijtw}$ were generated as described above. For each category (identified by the values of $n$, $m$, $\alpha$ and the way of generating times $t_{req}^j$), we generated 10 instances for a total number of 80 instances.\\

All computational tests were performed on an Intel i5 CPU @ 3.0 GHz with 16 GB of RAM. We implemented our B\&P scheme in C++ programming language. The Restricted Master Problem in the B\&P was solved by CPLEX 12.9 with a relative gap set to 0 and the barrier (interior point) algorithm. We benchmarked the performances of our B\&P algorithm on the considered instances against the performances of CPLEX 12.9 launched on model $BBAP_{ILP}$. In the performance comparison the parameters of the solver were set to their default values. We considered a time limit of 300 seconds for both CPLEX 12.9 and the B\&P algorithm. The choice of the time limit was related to the use of the proposed approach in daily operations, which usually require the computation of schedules in short running times.
\subsection{Results}
We report the computational results for the instances from Copenhagen airport in Table~\ref{tab:CPHinst}. In the considered operational days, there were time slots, such as the late night slots, with very few flights. Since the corresponding instances were trivial to solve, we present only the results for the instances with at least 20 flights landed at the airport in the associated time slot. In Table~\ref{tab:CPHinst}, we report, for each day, the minimum and maximum number of flights $n$ in a time slot, the number of belts $m$, the value of $\alpha$ considered for profit generation. Each day has 8 instances but the third day for which 7 instances were considered. The table reports the performance of CPLEX 12.9 launched on model $BBAP_{ILP}$ in terms of average computational time (column \textit{time}), average percentage gap $(\frac{UB^\prime}{LB^\prime} - 1)\cdot100$ between the best upper bound obtained ($UB^\prime$) and the best solution computed ($LB^\prime$) within the time limit (column \textit{gap (\%)}), the number of instances solved to optimality within the time limit (column \textit{\# opt}). For the B\&P algorithm, the table also reports the average number of nodes explored in the search tree (column \textit{\# nodes}). The average values consider also the instances where the solution methods reach the time limit.

\begin{table}[H]
	\centering
	\scalebox{0.8}{
\begin{tabular}{|l|r|r|*{1}{c|}*{3}{c|}*{4}{c|}}
  \hline
 &  &  & &\multicolumn{3}{|c|}{CPLEX 12.9} & \multicolumn{4}{|c|}{B\&P}  \\ \hline
 day & $n$	&	$m$ & $\alpha$ & time & gap (\%) & \# opt  & time &  gap (\%) & \# opt & \# nodes \\	\hline					
 1 & [20,35] & 5 & 0.5 & 5.83   & 0.00 & 8/8 & 12.35 & 0.00 & 8/8 & 381.00  \\
 &  &  & 0.8 & 6.24   & 0.00 & 8/8 & 11.30 & 0.00 & 8/8 & 361.00  \\ \hline		
2 & [20,43] & 5 & 0.5 & 46.32  & 0.02 & 7/8 & 16.63 & 0.00 & 8/8 & 401.00  \\
 &  &  & 0.8 & 49.79  & 0.04 & 7/8 & 13.97 & 0.00 & 8/8 & 320.38  \\ \hline		
3 & [25,41] & 5 & 0.5 & 173.48 & 1.66 & 3/7 & 50.65 & 0.02 & 6/7 & 547.14  \\
 & &  & 0.8 & 176.18 & 1.51 & 3/7 & 48.70 & 0.00 & 6/7 & 396.71  \\ \hline		
4 & [23,41] & 5 & 0.5 & 226.08 & 1.25 & 2/8 & 3.55  & 0.00 & 8/8 & 67.25   \\
 &  &  & 0.8 & 226.08 & 1.10 & 2/8 & 55.88 & 0.03 & 7/8 & 1018.50 \\ \hline		
5 & [22,44] & 5 & 0.5 & 152.24 & 0.78 & 4/8 & 7.18  & 0.00 & 8/8 & 84.75   \\
 & &  & 0.8 & 152.32 & 0.74 & 4/8 & 40.30 & 0.00 & 7/8 & 272.88  \\ \hline		
\end{tabular}}
		\caption{Real-life BBAP instances from Copenhagen airport.}
		\label{tab:CPHinst}
\end{table}
Our B\&P algorithm managed to solve to optimality 74 out of 78 instances with limited computational times. Notice that the percentage gaps were very small in the instances where the time limit was reached. The number of explored nodes along the branch operations was also reasonably limited. The proposed approach outperformed the solver CPLEX 12.9 that solved to optimality 48 instances. The solver had smaller computational times only in the instances of day 1.  \\

The same trend on the performance emerged in the results reported in Table~\ref{tab:RANDinst} for the randomly generated instances. The table has the same entries as those of Table~\ref{tab:CPHinst} except for the first column. In this column, we report the range of values of the requested times $t^j_{req}$. 
\begin{table}[H]
	\centering
	\scalebox{0.8}{
\begin{tabular}{|l|r|r|*{1}{c|}*{3}{c|}*{4}{c|}}
  \hline
 &  &  & &\multicolumn{3}{|c|}{CPLEX 12.9} & \multicolumn{4}{|c|}{B\&P}  \\ \hline
 $t_{req}\in $& $n$	&	$m$ & $\alpha$ & time & gap (\%) & \# opt  & time &  gap (\%) & \# opt & \# nodes \\	\hline							$[0, \frac{1}{2}t_{max}]$            & 30        & 5  & 0.5 & 300.00 & 10.74 & 0/10 & 100.74 & 0.00 & 9/10 & 1347.00 \\ 
			&         &   & 0.8 & 300.00 & 4.70  & 0/10 & 120.91 & 0.01 & 7/10 & 1531.90 \\ \hline
$[0, \frac{3}{4}t_{max}]$	  &30        & 5  & 0.5 & 300.00 & 4.16  & 0/10 & 116.71 & 0.03 & 7/10 & 1456.00 \\ 
	 &         &   & 0.8 & 300.00 & 1.79  & 0/10 & 77.82  & 0.00 & 8/10 & 996.80  \\ \hline
$[0, \frac{1}{2}t_{max}]$    &50        & 10 & 0.5 & 300.00 & 7.62  & 0/10 & 187.85 & 0.01 & 5/10 & 1463.30 \\
    &        &  & 0.8 & 300.00 & 4.77  & 0/10 & 230.31 & 0.00 & 4/10 & 1901.00 \\  \hline
$[0, \frac{3}{4}t_{max}]$ & 50        & 10 & 0.5 & 285.72 & 1.34  & 1/10 & 100.86 & 0.00 & 7/10 & 1106.50 \\ 
&         &  & 0.8 & 257.42 & 0.67  & 2/10 & 63.98  & 0.00 & 9/10 & 666.60  \\ \hline
\end{tabular}}
		\caption{Randomly generated BBAP instances.} 
		\label{tab:RANDinst}
\end{table}
The generated instances turned out to be more challenging to solve. This could reasonably be due to more narrow distributions of the delivery requested times and to the presence of belts with different productivity. Our B\&P algorithm solved to optimality 56 out of 80 instances. Still, the percentage gaps were significantly small and the approach strongly outperformed CPLEX 12.9. The solver was capable of solving to optimality 3 instances only. We also remark that our approach always provided better feasible solutions and upper bounds than CPLEX 12.9 in all instances where the time limit was reached. \\

Finally, we notice that the proposed B\&P algorithm 
on average, spent about the 90\% of the overall computational time in each tested instance
for solving the restricted master problems. 
This highlights the effectiveness of the dynamic programming algorithm in solving the pricing problems. Besides, the number of columns generated in the root node of the search tree was large enough (about 1068 on average) to allow a quick computation of feasible solutions either in the root node or during the exploration of the subsequent nodes.

\section{Conclusions}
\label{Conc}
We have presented a B\&P scheme for an optimal assignment of flights to baggage belts in the baggage reclaim area. The assignment ensures that for each belt, only one flight is serviced at each time. The approach takes care of a number of business and fairness constraints, avoiding congestions, and ensuring a good passenger flow. Robustness of the solutions is achieved by matching the delivery time with the expected arrival time of passengers, and by adding buffer time between two flights on the same belt. Computational experiments, based on real data from Copenhagen airport and on randomly generated instances, show that the proposed algorithm is effective in delivering high quality solutions in limited computational times, making it possible its use in daily operations in medium-sized and large airports. In future research, it would be interesting to extend the proposed algorithm to similar real-life applications and to further investigate the Position Dependant Knapsack Problem from both a theoretical and practical point of view.

\section{Acknowledgements}

The authors would like to thank Copenhagen Airport for having provided
data and insightful comments for this project.

\bibliographystyle{plain}
\bibliography{literature}

\begin{thebibliography}{10}

\bibitem{Barth2013}
T.~Barth.
\newblock Optimal assignment of incoming flights to baggage carousels at
  airports.
\newblock Technical Report Report 5, DTU Management Engineering, 2013.

\bibitem{Barth2013thesis}
T.~Barth.
\newblock {\em Optimization of Baggage Handling at Airports}.
\newblock PhD thesis, Department of Management Engineering, Technical
  University of Denmark (DTU), 2013.

\bibitem{Barth2012a}
T.~Barth and F.~B\"ockmann.
\newblock Baggage carousel assignment at airports: {M}odel and case study.
\newblock In {\em International Conference for Airport Operations Management in
  Munich}, pages 27--30, 2012.

\bibitem{Bor13a}
{G.M.R.} Borille and {A.R.} Correia.
\newblock Determining factors in airport baggage claim level of service.
\newblock {\em Internat. J. Aviation Management}, 2:66--79, 2013.

\bibitem{Bor13b}
{G.M.R.} Borille and {A.R.} Correia.
\newblock A method for evaluating the level of service arrival components at
  airports.
\newblock {\em Air Transport Management}, 27:5--10, 2013.

\bibitem{Delonge2012}
F.~Delonge.
\newblock Balancing load distribution on baggage belts at airports.
\newblock In {\em Operations Research Proceedings, International Annual
  Conference of the German Operations Research Society in Hannover}, pages
  499--505. Springer, 2012.

\bibitem{DeDeSo05}
G.~Desaulniers, J.~Desrosiers, and M.M. Solomon.
\newblock {\em Column generation}.
\newblock Springer US, 2005.

\bibitem{Frey2015}
M.~Frey.
\newblock {\em Models and Methods for Optimizing Baggage Handling at Airports}.
\newblock TUM-Bibliothek, 2015.
\newblock PhD thesis.

\bibitem{Frey2012}
M.~Frey, F.~Kiermaier, and R.~Kolisch.
\newblock Assignment of local baggage streams to claiming carousels at
  airports.
\newblock In {\em International Conference for Airport Operations Management in
  Munich}, pages 28--34, 2012.

\bibitem{Frey2017}
M.~Frey, F.~Kiermaier, and R.~Kolisch.
\newblock Optimizing inbound baggage handling at airports.
\newblock {\em Transportation Science}, 51:1210--1225, 2017.

\bibitem{KePfPi04}
H.~Kellerer, U.~Pferschy, and D.~Pisinger.
\newblock {\em Knapsack Problems}.
\newblock Springer, 2004.

\bibitem{MarTot90}
S.~Martello and P.~Toth.
\newblock {\em Knapsack Problems: Algorithms and Computer Implementations}.
\newblock Wiley, 1990.

\bibitem{Pis2020}
D.~Pisinger and {S. \'{i} H.} Rude.
\newblock Advanced algorithms for improved baggage handling.
\newblock {\em Journal of Airport Management}, 14, 2020.

\end{thebibliography}

\end{document}